\documentclass[%
superscriptaddress,
preprint,
showpacs,
aps,
prl,
]{revtex4-1}

\usepackage{graphicx}
\usepackage{dcolumn}
\usepackage{bm}

\begin{document}


\title{Link between K-absorption edges and thermodynamic properties of warm-dense plasmas established by improved first-principles method}
\author{Shen Zhang}
\affiliation{HEDPS, Center for Applied Physics and Technology, Peking University, Beijing 100871, China}
\affiliation{College of Engineering,  Peking University, Beijing 100871, China}
\author{Shijun Zhao}
\affiliation{HEDPS, Center for Applied Physics and Technology, Peking University, Beijing 100871, China}
\affiliation{College of Engineering,  Peking University, Beijing 100871, China}
\author{Wei Kang}%
\email[]{weikang@pku.edu.cn, xthe@pku.edu.cn}
\affiliation{HEDPS, Center for Applied Physics and Technology, Peking University, Beijing 100871, China}
\affiliation{College of Engineering,  Peking University, Beijing 100871, China}
\author{Ping Zhang}
\affiliation{HEDPS, Center for Applied Physics and Technology, Peking University, Beijing 100871, China}
\affiliation{ LCP, Institute of Applied Physics and Computational Mathematics, Beijing 100088, P.R. China}
\author{Xian-Tu He}
\email[]{weikang@pku.edu.cn, xthe@pku.edu.cn}
\affiliation{HEDPS, Center for Applied Physics and Technology, Peking University, Beijing 100871, China}
\affiliation{ Institute of Applied Physics and Computational Mathematics, Beijing 100088, P.R. China}

\date{\today}

\begin{abstract}
A precise calculation that translates shifts of X-ray K-absorption edges to variations of thermodynamic properties allows quantitative characterization of interior thermodynamic properties of warm dense plasmas by X-ray absorption techniques, which provides essential information for inertial confinement fusion and other astrophysical applications. We show that this interpretation can be achieved through an improved first-principles method.  Our calculation shows that the shift of K-edges exhibits selective sensitivity to thermal parameters and thus would be a suitable temperature index to warm dense plasmas. We also show with a simple model that the shift of K-edges can be used to detect inhomogeneity inside warm dense plasmas when combined with other experimental tools.


\end{abstract}

\maketitle
\thispagestyle{headings}


Warm dense matter (WDM) generally refers to a state of matter between solids and ideal plasmas. A typical WDM material usually has a density comparable to solids, and a temperature from several eV to tens of eV, \cite{barbrel2009} which are conditions fuel materials experience in the early stages of  inertial confinement fusion. \cite{atzeni2004} In addition, WDM broadly exists in various astronomical objects such as giant planets and brown dwarfs, \cite{Guillot1999} as well as in the core part of the earth. \cite{Huser2005}  Understanding the property of WDM is thus of particular interest to the investigation of these systems.  

With its large penetration depth and high resolution in time and space, \cite{yaakobi2008, levy2010, zhao2013} X-ray absorption is ideal for diagnosing interior properties of WDM, where X-ray absorption measurements provide information on electronic structures. Variations in thermodynamic properties are obtained by detecting induced changes in electronic structures.  
The position of K-absorption edge (K-edge) is defined by the transition between a K-shell electronic state and the lowest unoccupied electronic state. Since Bradley {\it et al.} \cite{bradley1987},  much effort has been spent trying to use the shift of K-edge to quantitatively characterize thermodynamic properties in a region well beneath the surface of WDM, which is of great interest yet not well understood.
The effectiveness of this approach depends  not only on accurate measurement of the X-ray absorption spectra, but also on the precision of calculations translating the shifts of K-edge energies into variations in thermal states.

Recent years have witnessed a substantial improvement in X-ray diagnostic techniques. \cite{levy2010, zhao2013, yaakobi2008} By X-ray absorption techniques, the K-edge  can now be determined with a temporal resolution less than 10 ps. \cite{levy2010} Theoretical methods based on first-principles molecular dynamics (FPMD), i.e., a combination of density functional theory (DFT) for electrons and classical molecular dynamics for ions,  have been established as effective in calculating thermodynamic properties for a variety of materials in their warm dense states. \cite{desjarlais2003, kietzmann2008, bonev2004, wang2011} Determining  K-edges of WDM using first-principles methods is more complicated. Unlike K-edge calculation for a crystalline structure, \cite{taillefumier2002, *gougoussis2009} where only a limited number of ions in a primitive cell have to be considered due to translational symmetry,  WDM K-edge calculation involves a large number of ions. Moreover, substantial influence of core electrons has to be taken into account properly.  These two factors, if not well handled, could cause unpredictable computational costs. The challenge is to find an appropriate treatment of core electrons in a system of a large number of ions to keep computational costs within the limit of current computational resources while maintaining the theoretical accuracy required. The serial work of Mazevet and  Z\'{e}rah, Recoules and  Mazevet, as well as Benuzzi-Mounaox {\it et al.} \cite{mazevet2008, recoules2009, benuzzi2011} on warm dense aluminum (Al) lays the foundation for accurate calculation of WDM K-edges.   By simplifying the treatment of core electrons using a pseudopotential method  within the framework of FPMD, they were able to obtain K-edges close to those measured below a density of 5 g/cm$^3$ along the principal Hugoniot of Al in shock experiments. However, when further compressed, their calculation \cite{mazevet2008, recoules2009, benuzzi2011} generally overestimates the magnitude of  K-edge shifts by more than 30\% (which is far beyond the error bars of experimental data), and the overestimation tends to increase with the Hugoniot compression. Essential improvement to the calculation  have to be made for quantitative characterization of thermal states of WDM.

In this work, we provide an improved FPMD calculation of K-edges for an extensively studied WDM material of shock-generated warm dense Al. \cite{desjarlais2002, zastrau2012, vinko2012, vinko2014, mazevet2008, recoules2009, benuzzi2011, hall1998, dasilva1989} 
The calculated K-edges display an excellent agreement with recent experimental data, as long as both K- and L-shell core electrons of Al are properly described. This allows a reliable translation between shifts in K-edge energies and variations in thermodynamic properties inside WDM.  Our results also reveal that the shift of the K-edge is more sensitive to the change of temperature than to the change of density, which indicates that the K-edge shift is a good index for interior temperature of WDM. In addition, the calculation suggest that when combined with other temperature measuring techniques, e.g., streaked optical pyrometers (SOP), \cite{miller2007, *zhang2014} the shift of the K-edge can be use to detect inhomogeneity inside WDM, which could provide further insights into the interior of WDM.

\begin{figure}
\centering
\includegraphics[scale=0.32]{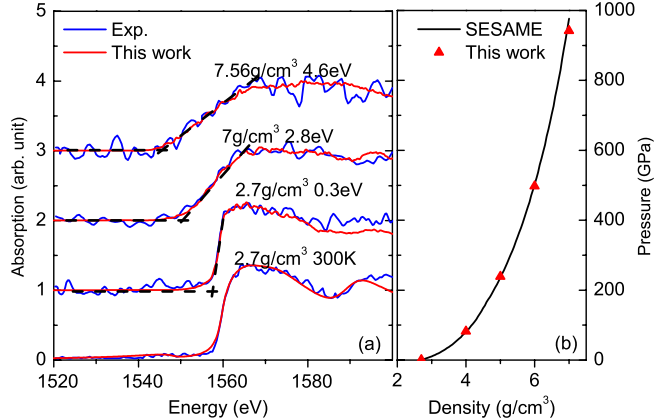}
\caption{\label{fig1} (a) X-ray absorption structures (XAS) of K shell electrons calculated (smooth curves) at selected thermal states of Al, compared with experimental measurements (undulating curves).  the XAS are presented in an arbitrary unit and shifted vertically with equal space for different thermal parameters. The slope of K-edges and corresponding abscissa are shown by dashed lines.  (b) Pressures calculated (triangles) along the principal Hugoniot. The solid line is the principal Hugoniot derived from the SESAME 3700 equation-of-state table, \cite{sesame} which is an accurate representation to the experimental Hugoniot of Al. }
\end{figure}

Our calculation consists of three consecutive steps: (i) Atomic trajectories are generated at given thermal states of warm dense Al, using the FPMD method together with an appropriately designed pseudopotential including both M- and L-shell electrons, which precisely accounts for electronic structures and ion-ion interactions under high pressure but still at a reasonable computational cost. The inclusion of L-shell electrons is revealed to be one of the crucial factors for getting accurate K-edge energies. It contributes more than 2/3 of the improvement, especially at high temperature.  (ii) Averaged X-ray absorption spectra (XAS) are calculated on atomic configurations uniformly sampled along the generated trajectory, and the K-edge position $E_{K,\mu}$ with respect to the chemical potential $\mu$ is then determined directly from the XAS as the intersection of its slope to the abscissa, as illustrated in Fig.~\ref{fig1}(a). (iii) The energy of 1$s$ states ($E_{1s}$) with respect to $\mu$ is determined by an all-electron DFT calculation on selected atomic configurations generated in the first step, which is the most challenge part of our calculations. In order to account for the temperature effect on $1s$ states with enough accuracy, more than 400 electrons have to be explicitly included in the calculation with a spatial resolution less than 0.2 Bohr for wave functions. It should be noted that the temperature effect has a substantial contribution to $E_{1s}$, but was not considered in the previous calculations. \cite{mazevet2008, recoules2009, benuzzi2011} Consequently, the K-edge energy $E_K$ is determined as the difference between $E_{1s}$ and $E_{k,\mu}$. 

Our calculations are carried out using the \texttt{Quantum Espresso} package \cite{pwscf}. The XAS are calculated using the \texttt{XSpectra} program \cite{taillefumier2002, *gougoussis2009} included in the package with minor modification to describe high-temperature electron distribution of WDM. A Perdew-Burke-Ernzerhof (PBE) type of exchange-correlation functional\cite{pbe} is used all through the work.

The first two steps of our calculation are similar to those employed in the previous studies, \cite{mazevet2008, recoules2009, benuzzi2011} but with a home-made pseudopotential including both M-shell and L-shell electrons (i.e., 2s$^2$2p$^6$3s$^2$3p$^1$) as valence electrons, which essentially improves the accuracy of electronic structure and ion-ion interaction. In the calculation, we adopt a plane-wave type FPMD together with the Born-Oppenheimer approximation, as implemented in the \texttt{Quantum Espresso} package. The pseudopotential takes the ultrasoft form \cite{uspp} with a core cut-off radius of 1.4 Bohr so that a plane-wave cutoff energy of 30 Ry and a shifted 2$\times$2$\times$2 k-point mesh can be used to further reduce computational costs. The atomic trajectories are generated in a canonical system, i.e., a system of constant NVT, consisting of 32 Al atoms in a cubic box with periodic boundary conditions assumed. A time step of 1 fs is used, and  atomic configurations in the last 1 ps are kept for the XAS calculation after the system evolves for more than 1 ps.  
The XAS   is averaged  on 8 snapshots uniformly picked from the trajectory of the last 1 ps. For each snapshot, the XAS is calculated following the established method in Ref.~\onlinecite{gougoussis2009}, which approximately includes electron-hole interactions by putting a hole state in the K-shell via a GIPAW pseudopotential. \cite{pickard2001} A shifted 4$\times$4$\times$4 k-point mesh and 400 electronic states  are used in the XAS calculation together with a plane-wave cutoff of 50 Ry to guarantee its accuracy up to 40 eV above K-edges.  As shown in Fig.~\ref{fig1}(a), the position of K-edge is measured directly from the XAS as the intersection of the K-edge slope to the abscissa.

$E_{1s}$ is determined by an all-electron plane-wave DFT calculation from the configurations used in the XAS calculation. Benchmark calculations on atomic Al show that, using a PAW pseudopotential \cite{blochl1994} together with a plane-wave cutoff of 400 Ry and a core radius cutoff of 0.15 Bohr, the energy of $1s$ state can be converged within 0.5 eV ($<0.1\%$) to the reference result obtained by any atomic all-electron code used to generate pseudopotentials.  To compare with experimental results directly, a constant energy shift of 63.0 eV is added to K-edge energies in order to compensate the underestimation to $E_{1s}$ caused by the DFT method itself.
\begin{figure}
\centering
\includegraphics[scale=0.4]{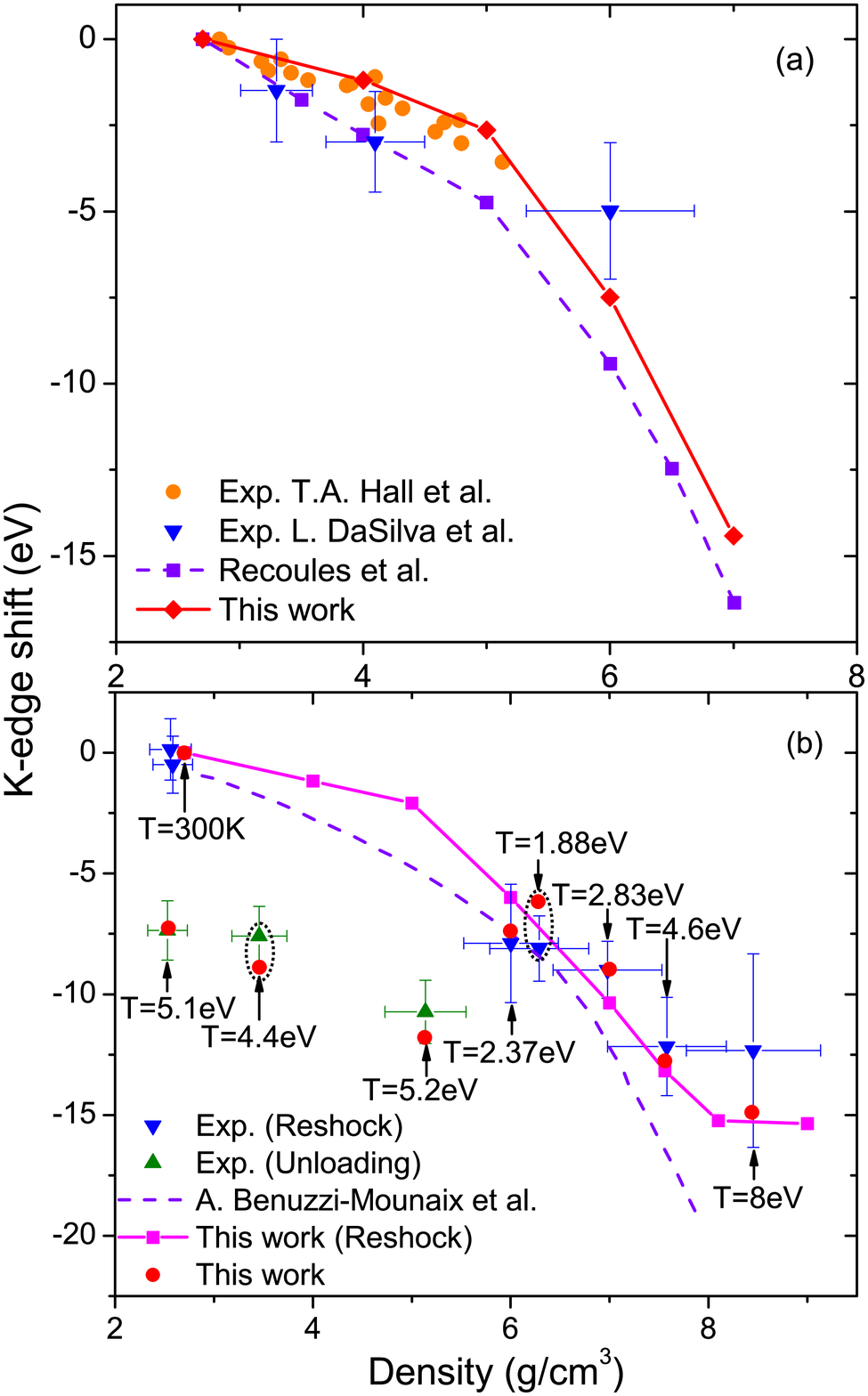} 
\caption{\label{fig2} Calculated K-edge shifts of Al referring to its uncompressed solid state, $\rho$=2.7g/cm$^3$ and  T=300K, compared with experimental measurements and previous calculations of similar methods. \cite{mazevet2008, recoules2009, benuzzi2011}  (a) K-edges calculated along the principal Hugoniot. Experimental data are taken from Hall {\it et al.} \cite{hall1998} and DaSilva {\it et al.} \cite{dasilva1989} The previous calculation is taken from Recoules {\it et al.} \cite{recoules2009} (b) K-edges under reshocked and unloading conditions, calculated with two subtly different sets of thermal parameters. Solid dots are calculated with thermal parameters measured by experiments, which are explicitly indicated in the figure. \cite{benuzzi2011} Solid curve with squares is calculated with parameters derived from a hydrodynamic simulation for the reshock condition. \cite{benuzzi2011} The two points slightly outside the experimental error bar are marked with doted circles. Calculations from Benuzzi-Mounaox {\it et al.} \cite{benuzzi2011} are also displayed for comparison. }
\end{figure}

Fig.~\ref{fig1} displays a comparison between our results  and some available experimental measurements. It gives an estimation to the confidence of our method in reproducing thermodynamic properties and electronic structures of warm dense Al. 
Selected XAS for typical thermal conditions are displayed in Fig.~\ref{fig1}(a) as smooth curves.  As a comparison, experimental XAS (undulating curves) for the same thermal parameters are also displayed. Close match between these two XAS suggests that the $E_{K,\mu}$ can be determined numerically within  $\sim$1 eV to experimental values. 
Fig.~\ref{fig1}(b) shows calculated pressure of Al along the principal Hugoniot. Also displayed is the Hugoniot derived from the SESAME 3700  equation-of-state table, \cite{sesame} which gives an accurate account for the experimental measurements of shocked Al. \cite{celliers2005} Our calculation agrees well with the experiments with an overall deviation less than 2\%.

\begin{figure}
\centering
\includegraphics[scale=0.35]{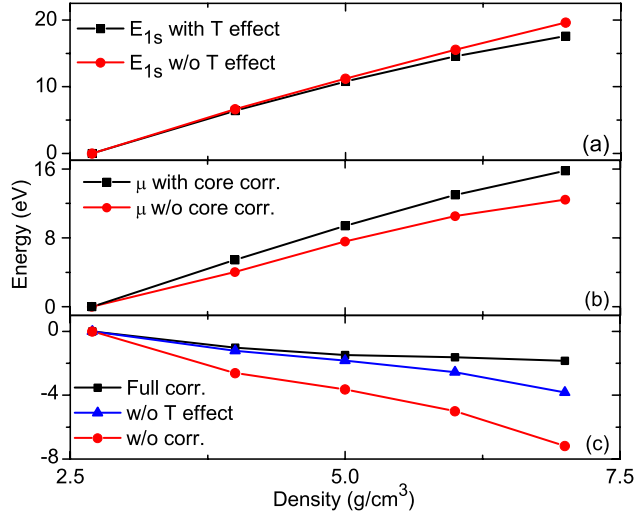} 
\caption{\label{fig3}Decomposition of contributions to the improvement of  K-edge calculation along the principal Hugoniot of Al,  referring to the standard solid state of Al. (a) Improvement to the calculation of E$_{1s}$ caused by temperature effect. (b) Improvement to the calculation of $\mu$ as a result of including L shell electrons explicitly in generating atomic trajectories. (c) Calculated energy of $\mu-E_{1s}$ with both corrections in (a) and (b), compared with the result before improvements taken into account.  An intermediate result (displayed as solid curve with triangles) including corrections to $\mu$ is displayed to visualize the fraction of contribution for each correction.}
\end{figure}

In Fig.~\ref{fig2}, we present calculated K-edge shifts of Al referring to its uncompressed state $\rho=$ 2.7 g/cm$^3$ and $T=$ 300K. Both calculated and experimentally measured K-edge shifts along the principal Hugiont are displayed in Fig.~\ref{fig2}(a).  The calculated results, shown as solid curves with diamonds, well reproduce the experimental results of Hall {\it et al.}, displayed as solid dots in Fig.~\ref{fig2}(a). \cite{hall1998} Earlier experimental results of DaSilva {\it et al.}, however, \cite{dasilva1989} exhibit an observable deviation $\sim$3 eV from our calculation and Hall {\it et al.}'s measurements at $\rho$ = 6.0 g/cm$^3$, as the result of low temporal resolution and insufficient characterization to plasma states in the earlier experiments. \cite{benuzzi2011}
Fig.~\ref{fig2}(b) shows that K-edge shifts under reshocked and unloading conditions can also be well described. 
The majority of our results are well located inside the experimental error bars, except two of them (highlighted by doted circles) having  slightly larger deviations. Since no systematic trends of these deviations are observed, they are probably caused by fluctuations in thermal parameters or by inhomogeneity of plasma states. Fig.~\ref{fig2}(b) also shows that K-edge shifts sensitively depend on thermal parameters of plasma states. The K-edge shifts calculated with instant thermal parameters have distinguishable differences from those (displayed as solid curves with squares) calculated with subtly different thermal parameters derived from a hydrodynamic code.\cite{benuzzi2011} The latter, representing K-edge shifts under ideal reshock conditions, approximately cross the center region of experimental data. 

The agreement between our calculation and experiments is attributed to a much improved estimation to $E_{K}$. 
There are two sources for the improvement. 
One is the temperature effect to $E_{1s}$, which was considered small and neglected in the previous calculations. \cite{mazevet2008, recoules2009, benuzzi2011} However, as we show in Fig~\ref{fig3}(a), this effect is substantial at a high compressing ratio when both $\rho$ and $T$ are high. The temperature effect goes into the correction indirectly. It first induces a spatial redistribution of ions and electrons at high temperature, which in turn causes a correction to the Coulomb potential energy part of $E_{1s}$.  
The other is the correction of L shell electrons included in the FPMD calculation, which gives a better account for wave functions at high density and  cause less blue shift in $\mu$, as illustrated by Fig.~\ref{fig3}(b). 
Fig.~\ref{fig3}(c) displays the net effect of these two contributions. It shows that the major contribution to the improvement comes from the correction to $\mu$, at all compressing ratios. Correction to $E_{1s}$ contributes less than 1/3 of the improvement, but increasing with further compression. Both corrections have significant contributions at a large compressing ratio.

\begin{figure}
\centering
\includegraphics[scale=0.39]{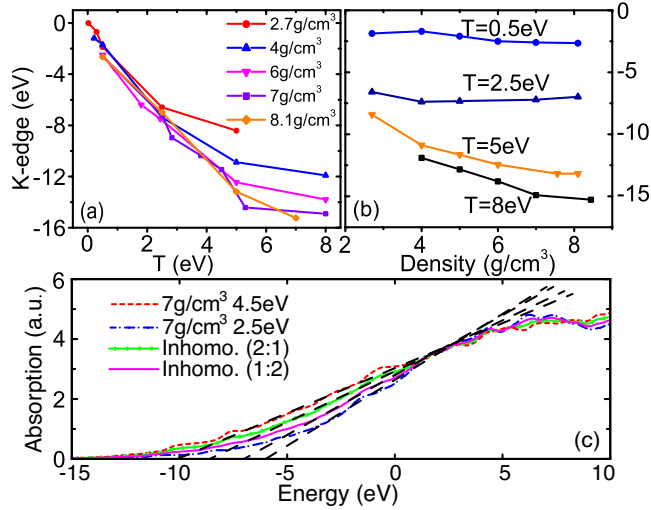}
\caption{\label{fig4} (a) K-edge shift as a function of temperature at fixed densities. (b) K-edge shift with respect to density calculated at constant temperatures. (c) XAS for  model inhomogeneous warm dense Al systems comprising two parts of the same density, but having different temperatures and volume ratios. XAS for the two homogeneous parts are also displayed as dashed curves for references. K-edge position is measured as the intersection of the slopes to the abscissa. The slopes are displayed as dashed lines in the figure.}
\end{figure}

To quantitatively characterize the relation between plasma states of Al and K-edge shifts, a systematic examination is presented in Fig.~\ref{fig4}(a) and (b), covering a variety states from solids to WDM.  K-edge shifts at different densities are displayed in Fig.~\ref{fig4}(a) as a function of temperature.  At low temperature, the K-edges decrease linearly at a slope of 2.9$\pm$0.2, which is almost independent to the variation of density.  When temperature further increases, a turning point occurs somewhere  between T = 2.5 eV and T = 5 eV, depending on the density of  WDM.  Fig.~\ref{fig4}(b) displays K-edges at different temperatures with respect to density. The flat shape of K-edges at low temperatures, as illustrated by the T = 0.5 eV and T = 2.5 eV curves, confirms the insensitivity of K-edge shifts to the variation of density. These results suggest that the K-edge has selective sensitivity to the variation of thermal parameters. In the parameter range investigated, K-edge is reasonably sensitive to the change of temperature and thus would be useful as a temperature index. Compared to the SOP technique, \cite{miller2007, zhang2014} which detects temperatures on the surface, K-edge shifts reflect the temperature inside WDM, which is a feature attractive to the study of bulk WDM.

Additionally,  we show with a simplified model that K-edge shifts, when combined with SOP, allow the probing of  inhomogeneity in bulk WDM.   XAS of model inhomogeneous warm dense Al systems are displayed in Fig.~\ref{fig4}(c). The model system comprises two homogeneous parts of the same density $\rho$ = 7.0 g/cm$^3$. The two homogeneous parts have different temperatures of T = 4.5 eV and T = 2.5 eV. They are put together with different mass ratio of 2:1 and 1:2 respectively. Dashed curves in Fig.~\ref{fig4}(c) represent XAS of the two homogeneous parts. Our calculation shows that the XAS of the inhomogeneous system is an mass-weighted average of these two homogeneous systems, and the K-edge of the inhomogeneous system is different from those of the two homogeneous systems. Since the shift of K-edges is less sensitive to the change of density in warm dense region, the density inhomogeneity is taken into account as the mass weight in the average.
A real inhomogeneous WDM system is composed of a large number of such small homogeneous parts along the path of X-ray. 
According to our calculation, the measured K-edge of a real inhomogeneous warm dense system is different from that calculated with the temperature measured by SOP at the surface.  Except for extreme cases where only density inhomogeneity exists, a deviation between these two K-edges thus indicates the appearance of inhomogeneity.   

In summary, we show that an accurate estimation to the K-edge shift of warm dense Al can be achieved by an improved first-principles calculation when the effect of core electrons are carefully taken into account. A calculation of such accuracy would open a new possibility for X-ray absorption technique to quantitatively characterize internal plasma states of WDM, which is of particular interests to a variety of fields including ICF, astrophysics, and geophysics as well.

This work is supported by the National Natural Science Foundation of China (Grant No. 11274019).

\bibliographystyle{apsrev4-1}
%

\end{document}